**Capacitance modeling of complex topographical silicon quantum dot structures**

H. L. Stalford[1,3], R. Young[1], E. P. Nordberg[1,2], James. E. Levy[1], Carlos Borras Pinilla[3], M. S. Carroll[1]

[1]Sandia National Laboratory

[2]University of Wisconsin-Madison

[3]University of Oklahoma

Abstract

Quantum dot (QD) lay-outs are becoming more complex as the technology is being applied to more complex multi-QD structures. This increase in complexity requires improved capacitance modeling both for design and accurate interpretation of QD properties from measurement. A combination of process simulation, electrostatic simulation, and computer assisted design (CAD) lay-out packages are used to develop a three dimensional (3D) classical capacitance model. The agreement of the classical model's capacitances is tested against two different, experimentally measured, topographically complex silicon QD geometries. Agreement with experiment, within 10-20%, is demonstrated for both structures when the details of the structure are transferred from the CAD to the model capturing the full 3D topography. Small uncertainty in device dimensions due to uncontrolled variation in processing, like layer thickness and gate size, are calculated to be sufficient to explain the disagreement. The sensitivity of the capacitances to small variations in the structure also highlights the limits of accuracy of capacitance models for QD analysis. We furthermore observe that a critical density, the metal insulator transition,[1] can be used as a good approximation of the metallic edge of the quantum dot when electron density in the dot is calculated directly with a semi-classical simulation.

1. **Introduction**

Electrostatically defined quantum dots (QD) are of increasing interest because of technological scaling (e.g., floating gate memory)[2-3], state-of-the-art standards, and revolutionary advances beyond CMOS (Complementary Metal Oxide Semiconductor) computing approaches such as cellular automata[4] and quantum computing[5]. Sophisticated geometries have been and are increasingly being experimentally examined for applications such as quantum dot cellular automata[4], charge pumps[6-7], single electron charge sensing[8-9], multiple QD coupling[10-11], circuit interactions with QDs[12] and proposed architectures for different kinds of computing[5,13-14]. This increasing sophistication in QD technology that involves many QDs coupled to neighboring devices such as charge sensors and external circuitry requires improved modeling capabilities both to design rapidly, assist in analyzing results (like dot size and position) as well as understand potential systematic parasitic effects such as disorder within the quantum dot and external cross-talk.

Quantum dot behavior is frequently dominated by classical capacitive properties of the structure[15]. Highly detailed models that incorporate calculations of the quantum mechanical aspects of the dot are critical for in depth understanding of the underlying single electron physics. However, a purely classical capacitance model is sufficient for many necessary aspects of QD and QD coupling to the environment. Furthermore, a less numerically intensive model than those required for full self-consistent Schroedinger-Poisson or technology CAD (TCAD) calculations are often necessary to tractably handle many coupled devices and the surrounding environment when it is desired to rapidly investigate numerous permutations of QD device geometries and their surrounding couplings.

In addition to guiding the design and subsequent iterations of QDs the capacitive coupling model will be extremely useful to engineers designing the future support electronics needed to control, manipulate, and measure QDs. The capacitance model can be translated into a SPICE[16] (Simulation Program with Integrated Circuit Emphasis, [SPICE]) circuit element to determine the effects that control and readout will have on the QD. Effects worth exploring include cross-talk between signal lines[17], feedback to the QD from readout, charge injection[18], and the effects of process variation establishing limits on the precision and accuracy of such capacitance models.

In this paper, we investigate the accuracy in calculated capacitances for topographically complex silicon QD structures. Experimental measurement of the quantum dot size is established through the capacitances of the external conductors to the quantum dot. A three dimensional (3D) capacitance matrix is calculated for two significantly different quantum dot geometries and compared to their respective experimental results published in the literature. A heuristic approach is examined in which the size of the quantum dot, the metallic sheet of quantum confined electron density, is estimated by the lithographic features of the structure and then slowly adjusted to fit measurement. This approach is compared with an approach in which the quantum dot size is estimated by directly simulating the electron density using semi-classical methods available in a commercially available TCAD simulator. Agreement with experiment, within 10-20%, is demonstrated for both approaches when the details of the full structure are transferred into the 3D solid-model. Uncertainty in the actual experimental structure due to small variations in layer thickness and gate sizes are more than sufficient to explain differences between the simulation and experimentally obtained capacitances. We present, furthermore, the observation that the size of the quantum dot, the metallic quantum confined region, can be estimated well using an electron density contour of $1.5 \times 10^{11}$ cm$^{-2}$ combined with the semi-classical TCAD calculation of the electron density. The metal-insulator transition for similar silicon MOS devices was measured recently[1] to be approximately $1.5 \times 10^{11}$ cm$^{-2}$. This model's utility is also demonstrated through its application in corroborating that the quantum dot size is defined by the lithographic gate features rather than disorder in the local potential. Disorder in the MOS system, non-uniform potential due to charged defects and traps, can lead to disorder assisted quantum dot formation. The sensitivity of this capacitance simulation to the size and location of the quantum dot, therefore, lends significant assistance in determining whether the quantum dot is formed with assistance of a local disorder potential or whether the system is relatively clean of disorder and determined by the intended physical gate structure.

## 2. Description of Approach

The following general approach was applied to a tunable lateral QD geometry[19] and a top gated nanowire geometry[20] both of which were enhancement mode MOS structures. A detailed description of the tunable lateral QD calculation will be described, while only the results of the gated nanowire calculation will be presented. The general approach to calculating the capacitances begins with constructing a meshed 3D model of a quantum device that is then used for simulating capacitances.

The steps for simulating capacitance matrix starts with developing an AutoCAD model, Fig. 1. Either the original AutoCAD model for the lithography is used or a more accurate representation is produced by altering the original AutoCAD to match SEM images of the fabricated nanostructures.

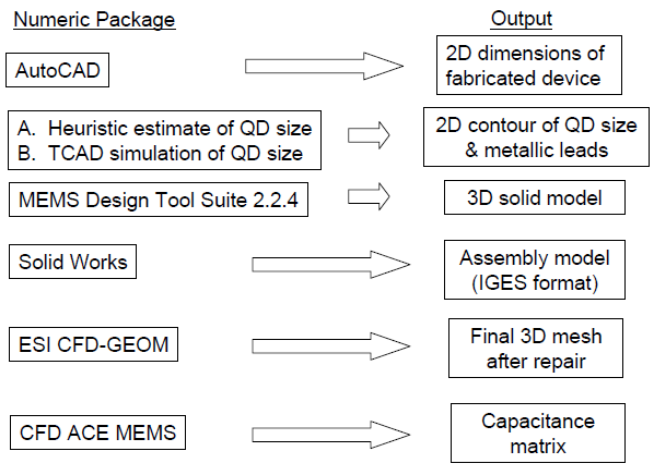

Fig. 1. Flow chart for simulating capacitance matrix

The conducting regions designated by the metallic regions in the AutoCAD model are augmented with the incorporation of regions that represent thin metallic sheets in the silicon where the formation of either a 2-dimensional electron (or hole) gas (2DEG) or quantum dot region are expected. Two approaches are used in defining the conductor elements of the 2DEG layer. One is to use a heuristic approach starting with the 2DEG position assumed to be located directly below the lithographic gate features that are biased to induce electrons or holes. A second approach, which is more computationally intensive, uses a commercially available semi-classical TCAD simulation package to calculate the electron density in limited regions to provide better local estimates of the metallic region. This provides a more accurate estimate of effects of the 3D topology and lateral depletion from neighbor gates used to control the lateral extent of the electrons (holes). For this case, a critical density is chosen to define the edge between metallic and insulating behavior. The metal-insulator transition (MIT) is defined at a density of $1.5 \times 10^{11}$ electrons/cm$^2$ in this work, which was a measured transition[1] at similar operating temperature for the devices examined in this work, T ~ 300 mK.

Several CAD packages are engaged to achieve a high fidelity of transfer of the 3D features of the device to the simulated capacitance model. A custom fabrication process file (e.g., what is used for e-beam lithography), compatible with AutoCAD software and a process simulator, MEMS Design Tool Suite 2.2.4[21], is used to render the essential features of the fabrication process. A 3D solid model is generated with this CAD software package. The 3D solid model is imported from AutoCAD into SolidWorks[22], [SolidWorks], where an assembly model is built and saved as an Initial Graphics Exchange Specification (IGES) file which, in turn, is then imported into ESI's CFD-GEOM[23]. Manual repairing operations (sometimes extensive on complex geometries) are needed at each stage to clean up missing or defective surfaces and erroneously appearing lines and points (so-called "dirty" geometry entities) imported from other CAD software. CFD-GEOM'e editor offers simple and easy-to-use tools to repair and modify such geometry faults. The 3D model with clean geometry is meshed in CFD-GEOM. The meshed 3D model from CFD-GEOM is used in CFD ACE+MEMS electrostatics solver[23], [ESI-Group], to compute the capacitance matrix. CFD-ACE+MEMS software provides a fully integrated environment for multiscale, multiphysics, high fidelity analysis of semiconductor device structures.

## 3. Modeling of Double Top Gated Quantum Dot & Double Gated Nanowire

Capacitances are calculated and compared to measurement for two quantum dot device structures with significantly different geometries. An open, double-top-gated lateral structure with significant topology[19], Fig. 2, is designated device 1 for the discussion in this paper and leads to a 3D model with 16 conductors. The device structure uses two levels of conducting gates. A top Al gate is biased positively to draw electrons into the silicon, and poly-silicon gates buried below the Al are used to locally deplete and confine the electrons in low dimensional geometries. The conducting gates are insulated from one another and the silicon using a SiO$_2$ gate oxide between the poly-Si and the Si combined with an atomic layer deposited (ALD) $Al_2O_3$ layer placed between the poly-Si and the Al. The simulated structures assume a conformal deposition of the ALD using the same thickness on sidewalls as on top of the flat regions. The layer thicknesses are nominally the same as those used in the experiment: $SiO_2$ gate oxide=35 nm; polysilicon=200 nm; atomic layer deposition (ALD) of $Al_2O_3$=60 nm; and Al top gate=300 nm.

The silicon substrate is modeled as a dielectric to a depth of 1000 nm. The actual geometry of the fabricated polysilicon gate structure is taken from a SEM (scanning electron microscope) image of a duplicate device, Fig. 3, after the polysilicon etch step. In developing an AutoCAD file for simulating capacitances, we started with the original AutoCAD geometry used in developing the mask for etching the polysilicon structures and modified it to match the SEM image. A good match to the fabricated polysilicon structures, Fig. 4, was achieved by enlarging the original AutoCAD poly-Si dimensions an additional 30 nm in all directions beyond their dimensions in the poly-Si etch mask.

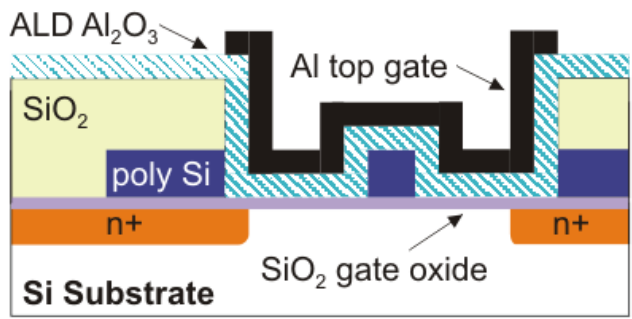
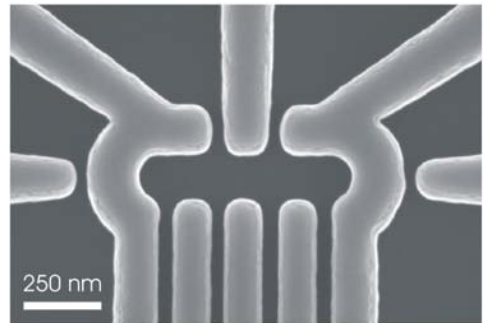

Fig. 2. Cross-sectional schematic of simulated silicon lateral quantum dot structure described in text and from reference 19. The cross section shows only one active poly-silicon depletion gate. The lay-out of the lateral poly-silicon depletion gate structure is shown in figure 3.

Fig. 3. Scanning electron microscope image of poly-silicon gate structures from lateral quantum dot structure described in text and in reference 19.

The polysilicon gate structures, Fig. 4, are denoted as conductors C8-C15 in the 3D simulation model. The Al top gate (i.e., accumulation gate), is designated as conductor C16. The top gate is used to induce regions of sufficient electron densities in the silicon at the Si-$SiO_2$ interface that are higher than the metal insulator transition (i.e., $1.5 \times 10^{11}$ electrons/$cm^2$) resulting in metal-like conductors. For the purpose of developing the 3D simulation model, the 2DEG regions are assumed to extend a depth up to 10 nanometers into the silicon from the oxide interface. As a preliminary estimate of the quantum dot island designated as C1 in Fig. 5, we assumed an oval shape with width 140 nm and length 680 nm. The actual size and shape of the island C1 depends on voltages placed on the accumulation and depletion gates C8-C16 and is modified later below accordingly for specific conditions. The boundaries of the polysilicon gates C8-C15, Fig. 4, were used as an offset reference in defining the boundaries of the 2DEG regions C2-C7, Fig. 5. A 5nm horizontal offset outside the gate boundaries was used. That is, from a top view, they appear to be offset horizontally by 5nm outside their reference. We emphasize that the size and shape of the metal-like conductors C1-C7 depend largely on sufficient electron density created by voltages placed on the gates C8-C16. The 2DEG conductors C1-C7 of the 3D solid model can be modified to match the experimentally observed capacitances for each voltage condition while the rest of the 3D solid model remains without change. This is the heuristic approach.

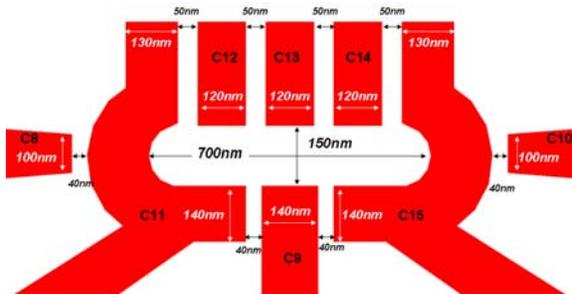

Fig. 4. Geometry of fabricated polysilicon gates C8-C15. Simulated dimensions are indicated on figure.

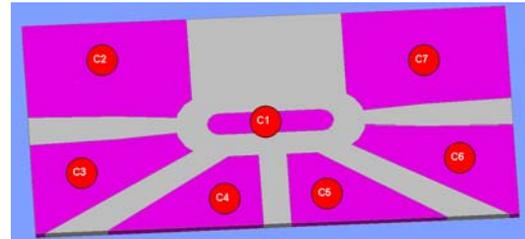

Fig. 5. Location of 2DEG, metal-like conductors C1-C7, assumed to have 10nm depth in the silicon substrate.

Three dimensional views of the simulation model are given in Figs. 6A-6D wherein the 1000 nm Si-substrate of the model has been removed.

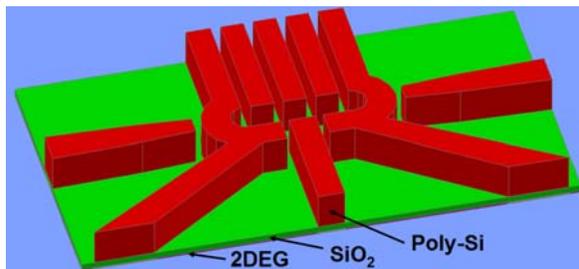

Fig. 6A. Three dimensional view of model with 2DEG, 35 nm $SiO_2$ gate oxide and poly-Si layers (1000 nm Si-substrate not shown).

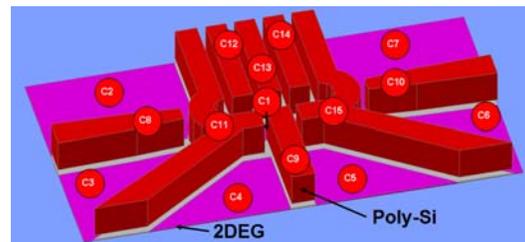

Fig. 6B. Three dimensional view of conductors C1-C15 with 35 nm $SiO_2$ gate oxide layer removed for viewing 2DEG

layer.

In Fig. 6A, the polysilicon structures C8-C15 are as deposited on top of the 35 nm SiO$_2$ gate oxide layer which covers the 2DEG layer as shown by its removal in Fig. 6B. The dielectric 60 nm Al$_2$O$_3$ layer, Fig. 6C, is a simulated conformal deposition over the polysilicon structures. The Al top gate C16 conductor, Fig. 6D, is sputtered over the Al$_2$O$_3$ layer and conforms to its shape. While the sputtered Al metal is not a conformal deposition, it can be treated as such for electrostatic simulation purposes since it is thick enough to coat vertical and horizontal surfaces of the Al$_2$O$_3$ layer.

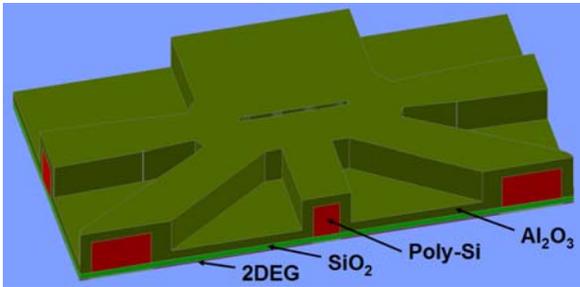

Fig. 6C. Three dimensional view of Al$_2$O$_3$ layer covering poly-Si, SiO$_2$ gate oxide and 2DEG layers. The 1000 nm Si-substrate is not shown.

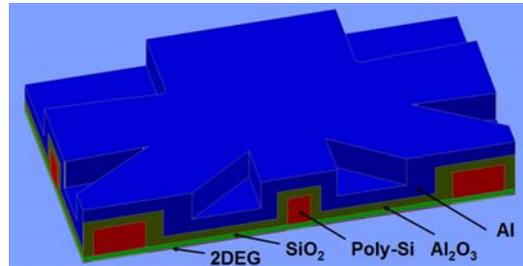

Fig. 6D. Three dimensional view of conductor Al top gate layer, C16, covering Al$_2$O$_3$, poly-Si, SiO$_2$ gate oxide and 2DEG layers. The 1000 nm Si-substrate is not shown.

Combining the 2DEG "conductors" C1-C7 together with the depletion gates C8-C15 and the Al accumulation top gate C16 makes a total of 16 conductors. Below we use 3D simulation to determine the capacitances between all the 16 conductors. In particular, we are interested in the capacitances between the gates C8-C16 and the quantum dot island C1. After the 3D solid model is developed, it is meshed in CFD-GEOM.

Various aspects of the meshed 3D solid model are presented in Figs. 7A-7D and include the following as shown in Fig. 7D, proceeding from the bottom to the top: 1000 nm silicon substrate, 10 nm 2DEG layer with conductors C1-C7, 35 nm SiO$_2$ gate oxide layer, 200 nm poly-Si depletion gate layer with conductors C8-C15, 60 nm ALD Al$_2$O$_3$ conformal layer, and 300 nm Al accumulation gate. In Fig. 7A, the faintly visible yellow lines define the boundaries of the 2-DEG Conductors C2-C7 and a solid yellow oval is superimposed to highlight the dot C1. The CFD GEOM meshed model for Device 1 used 3D-unstructured grid tetrahedral elements of the finite volume method (FVM) and had the following parameters: (a) The minimum cell size was 0.01 nm, (b) The maximum cell size allowed was 30 nm, and (c) The total number of unstructured domain tetrahedral elements was 2.2 million. The simulated and experimental capacitances for this structure are presented below in section 4.

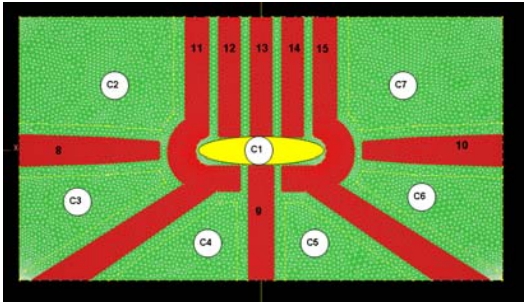

Fig. 7A. Meshed 10 nm depth 2DEG layers with conductors C1-C7.

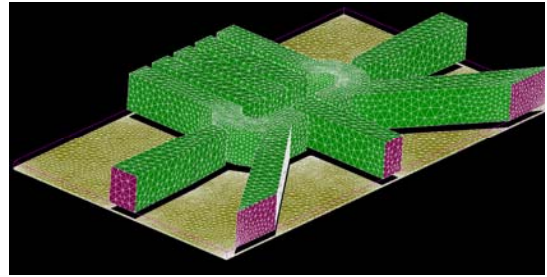

Fig. 7B. Meshed depletion gates C8-C15 and 2DEG 10 nm layer.

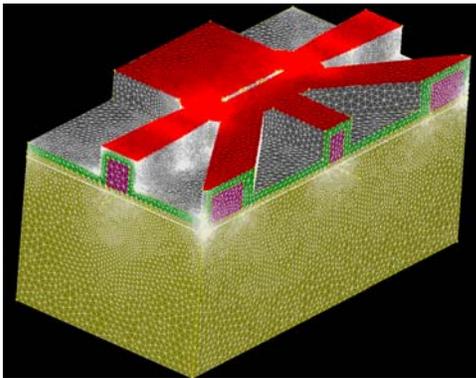

Fig. 7C. All meshed elements without Al top gate C16.

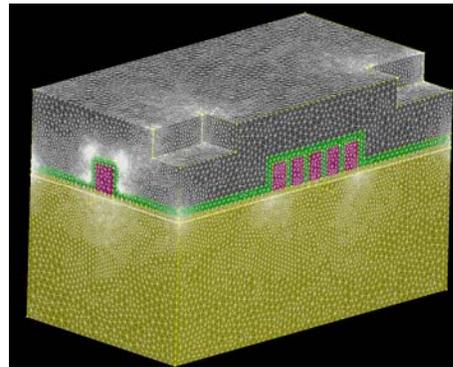

Fig. 7D. Meshed 3D solid model of quantum device.

The second quantum device modeled is a multi-gated silicon nanowire metal-oxide-semiconductor field-effect transistor (MOSFET)[20] for which the experimental capacitances between the gates and the resulting quantum dot have been characterized in detail. As shown in Fig. 8 with top view dimensions, it is an 8-conductor model with source and drain (C3 and C4), Si charge islands (C1 and C2), lower poly-Si gates (C5, C6 and C7) and upper poly-Si gate (C8). The source and drain conductors C3 and C4 are 400nm by 400nm areas with extensions along the nanowire as shown. The lower poly-Si gates C5 (LGS), C6 (LGC), and C7 (LGD) have 10 nm widths and 370 nm lengths. The upper poly-Si gate conductor C8 (UG) is a 600 nm by 600 nm area. The charge islands C1 and C2 have 20 nm widths. The lengths of the two charge islands depend on the length of the three barrier gaps in between the source and drain conductors and the two charge islands. We assume that the lengths of the three barrier gaps are equal and that they are positioned symmetrically as shown underneath the three lower poly-Si gates C5-C7.

3D aspects of the model are given in Figs. 9A-9F. The Si-nanowire runs between the source and drain, Figs. 9A and 9B, and has layer thickness (20 nm) and width (20 nm); this includes the two charge islands C1 and C2 and the three Si-barrier gaps.

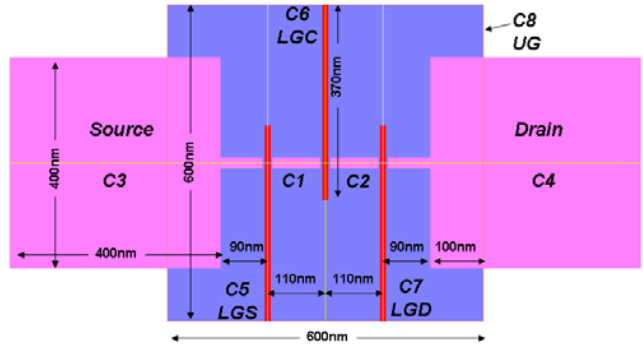

Fig. 8. Top view of basic model dimensions of nanowire device

The Si-nanowire layer resides on top of a buried $SiO_2$ layer with 400 nm thickness. The lower poly-Si gates C5-C7 in Fig. 9B have 10 nm widths. The thickness of the poly-Si gate layer is taken to be 30 nm, an estimation based on images of the device[20].

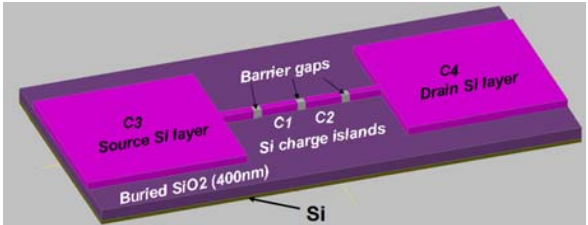

Fig. 9A. The 20nm by 20nm Si-nanowire resides on top of a 400nm buried SiO2 layer and Si substrate.

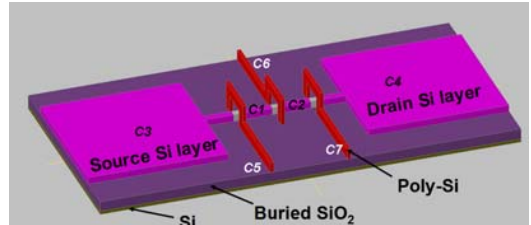

Fig. 9B. The lower poly-Si gates C5-C7 have 10nm widths (gate oxide not shown).

As shown in Fig. 9C, a 30 nm gate oxide $SiO_2$ covers the nanowire and electrically insulates the conductors C1-C4 from the lower poly-Si gates C5-C7. Another 30 nm thick $SiO_2$ layer is used to cover the lower poly-Si gates and insulate them from the upper poly-Si gate C8, Figs. 9D-9F. The air spaces on each end of the nanowire device are included as part of the 3D solid model, Figs. 9E and 9F; thus, the air spaces are included in the calculation of field lines for the entire model.

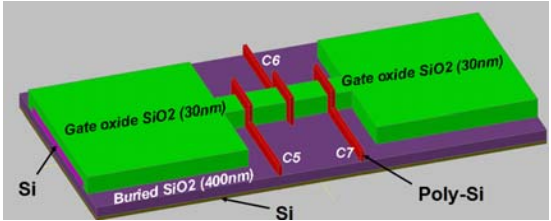
Fig. 9C. The nanowire conductors C1-C4 are covered by a 30 nm SiO2 gate oxide.

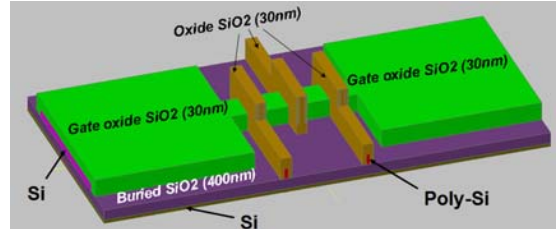
Fig. 9D. The lower poly-Si gates C5-C7 are covered by a 30 nm SiO2 gate oxide.

Examples of the meshed elements are given in Figs. 10A and 10B. As shown in Fig. 10A, the gates surround the etched wire on all three sides as shown by purple shading on the edge of the gray wire. Thus, the nanowire device is like a FinFET (i.e., Field Effect Transistor with vertical Si fin controlled by multigates).

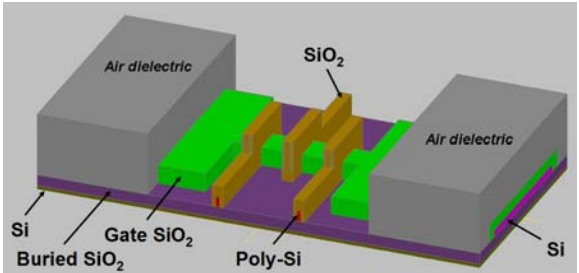
Fig. 9E. The 3D solid model provides for air dielectrics on each end.

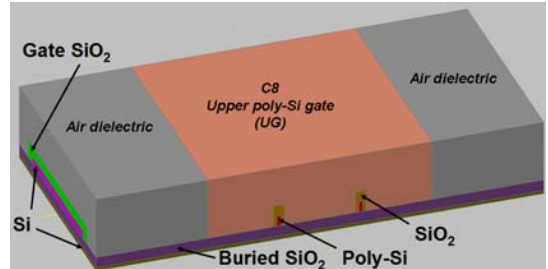
Fig. 9F. Complete 3D solid model of nanowire device with upper poly-Si gate.

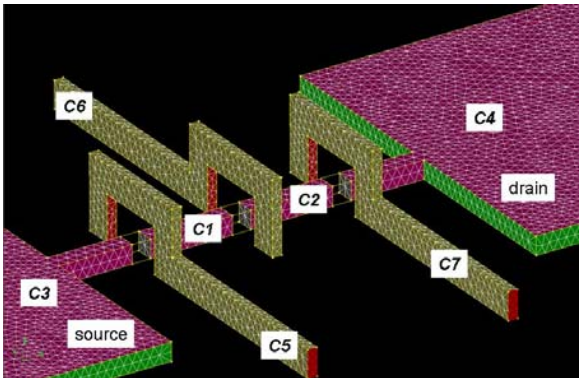
Fig. 10A. Nanowire model with lower poly-Si gates C5-C7, charge islands C1 and C2, and source and drain conductors C3 and C4 (upper gate C8 is not shown)

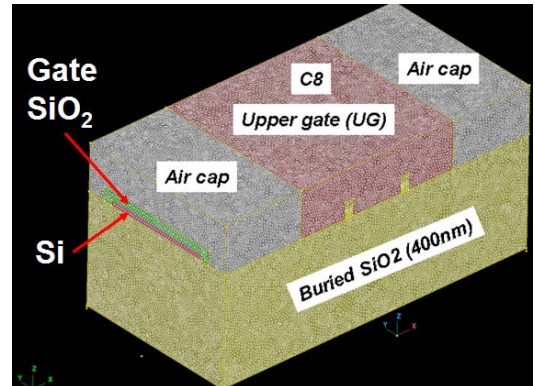
Fig. 10B. Complete 3D nanowire model with dielectrics (buried $SiO_2$, gate $SiO_2$, Si barriers, and air caps) and conductors C1-C8.

## 4. Capacitance Results/Comparison with Experimental Measurements

ESI's CFD-ACE+MEMS software with its advanced electrostatic solver is used in computing the capacitance matrix for quantum devices with 3D solid models. First, we illustrate this using the 16-conductor model described above in Figs 4-7. The dielectric permittivities for $Al_2O_3$, $SiO_2$, and the silicon substrate were taken to be 7.9, 3.9, and 11.9, respectively. Taking each conductor in turn, an applied potential of 1 volt is placed on the selected conductor and zero potential on the rest. Using the constraint of zero net charge, the electrostatics problem is solved for the net charge on each conductor. The nodal capacitance matrix equation [Q]=[C][V] which provides a relationship between charge, voltage, and capacitance is used in computing the capacitance matrix. The 16 by 16 capacitance matrix is symmetric with its diagonal terms positive to denote the self-capacitance of the matrix (i.e., with all conductors in place) and off diagonal terms negative to denote mutual capacitance. The sum of all elements in every column and every row is equal to zero, the result of charge conservation (i.e., zero net charge). The simulated capacitances between the conductors C8-C16 and the quantum dot island C1 are shown in Fig. 11. We note that the self-capacitance calculated here is not the same as the capacitance that would result from a charge moving from infinity to the isolated conductor without any metal around it. This term must be calculated separately and is necessary for calculating the charging energy of a quantum dot.

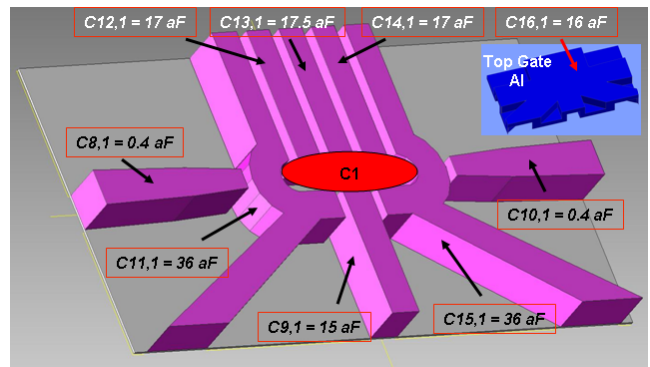

Fig. 11. Capacitances between conductors C8-C16 and the quantum dot C1 for the quantum device[19].

Stable, repeatable Coulomb blockade was observed in both devices. The period of the blockaded transport can be used to extract the capacitance between the gates and the quantum dot[15]. First, we simulated capacitances for the lateral device and compared them with the experimental values, Table 1. No fitting parameters were used. The 2DEG islands C2-C7 were assumed to have edges 5 nm from gates C8-C15. The 2DEG island C1 was assumed as an oval shape 680 nm by 140 nm, fitting inside an oval space 700 nm by 150 nm. Next, we compute the capacitance matrix for the 8-conductor nanowire model described above, Figs 8-10, using a fixed barrier gap of 7 nm, Table 2. Although not considered herein, the barrier gap between leads and islands could have been used as a fitting parameter since some of the capacitances (e.g., C3,1, C4,2) will depend on the value assumed for the barrier gap.

The simulated capacitances, taken from Table 2, are presented in the 4$^{th}$ column of Table 3 for comparison with previously published experimental results for the nanowire device. The experimental capacitances, first three columns of Table 3, were estimated from Coulomb blockade oscillations[20].

Table 1. Comparison of simulated and measured capacitances for lateral quantum device

| Gate | Model 1 | Measured |
|---|---|---|
| C9,1 | 15 aF | 20.8 aF |
| C11,1 | 36 aF | 42 aF |
| C12,1 | 17 aF | 14.7 aF |
| C13,1 | 17.5 aF | 11.2 aF |
| C14,1 | 17 aF | 14.1 aF |
| C15,1 | 36 aF | 42 aF |

Table 2. Simulated capacitance matrix for nanowire device in which capacitance values are in units of attofarads (**aF**).

| | | Island #1 | Island #2 | Source | Drain | LGS | LGC | LGD | UG |
|---|---|---|---|---|---|---|---|---|---|
| | | C1 | C2 | C3 | C4 | C5 | C6 | C7 | C8 |
| Island #1 | C1 | 30.5 | -7.7 | -8.1 | -0.23 | -2.3 | -2.3 | -0.06 | -9.7 |
| Island #2 | C2 | -7.7 | 30.5 | -0.23 | -8.1 | -0.06 | -2.3 | -2.3 | -9.7 |
| Source | C3 | -8.1 | -0.23 | 130 | -2.3 | -3.7 | -0.59 | -0.32 | -115 |
| Drain | C4 | -0.23 | -8.1 | -2.3 | 130 | -0.32 | -0.59 | -3.7 | -115 |
| LGS | C5 | -2.3 | -0.06 | -3.7 | -0.32 | 72 | -0.14 | -0.08 | -65 |
| LGC | C6 | -2.3 | -2.3 | -0.59 | -0.59 | -0.14 | 72 | -0.14 | -66 |
| LGD | C7 | -0.06 | -2.3 | -0.32 | -3.7 | -0.08 | -0.14 | 72 | -65 |
| UG | C8 | -9.7 | -9.7 | -115 | -115 | -65 | -66 | -65 | 445 |

Table 3. Comparison of simulated and measured capacitances for nanowire device

|  | Experimental Results[20] | | | Simulation |
|---|---|---|---|---|
|  | Device 1 Short Islands Source/Drain Capacitance (aF) | Device 2 Short Islands Source/Drain Capacitance (aF) | Device 3 Short Islands Source/Drain Capacitance (aF) | 3D Simulation Short Islands Source/Drain Capacitance (aF) |
| CUG | 10/11 | 11/11 | 10/11 | $C_{8,1}/C_{8,2}$ = 9.7/9.7 |
| CLGS | 2.8/0.09 | 2.3/0.12 | 2.9/0.07 | $C_{5,1}/C_{5,2}$ = 2.3/0.06 |
| CLGC | Not measured | 2.8/3.1 | 2.6/2.8 | $C_{6,1}/C_{6,2}$ = 2.3/2.3 |
| CLGD | 0.08/2.4 | 0.14/2.4 | 0.08/2.5 | $C_{7,1}/C_{7,2}$ = 0.06/2.3 |

Comparison of the modeling results with the measurements shows simulation to be within 10-20% in all cases, which represents good agreement especially for these complex topographies.

## 5. TCAD Refinement of Quantum Dot Geometry and Resulting Simulated Capacitances

In this section we describe a TCAD assisted approach to calculating the full capacitance matrix over all gates. We utilize TCAD to compute the electron density within a sub-section of the entire structure of the first device, Fig. 12. The $1.5 \times 10^{11}$ electrons/cm$^2$ density contour is used to define the metallic edge of the quantum dot for the capacitance modeling. The critical contour density is guided by previous measurements of the critical density at which 2DEGs in the MOS system reach the metal insulating transition[1], $\sim 1.5 \times 10^{11}$ electrons/cm$^2$. We then simulate the capacitances using this critical density-based geometry of the dot in the model. Since the size and shape of the quantum dot depends on the gate voltages, this approach is necessary in general partly due to the fact that the TCAD analysis is based on the voltages placed on the gates.

The 2DEG electron density contours and inversion layer depth in the silicon substrate are determined by performing 3D simulations and analysis using TCAD software Taurus Medici (previous name was Davinci) from Synopsys, Inc[24]. A symmetric 3D model of the right half of the structure was used in the TCAD simulations, Fig. 12, in which the depletion gate structures (i.e., C9, C11-C13) are shown as "rose".

The "blue" regions in Fig. 12 are where the top gate is at its lowest point, a separation of 95 nm from the 2DEG layer (i.e., the thickness of the $SiO_2$ gate oxide plus the thickness of the $Al_2O_3$). The contour enclosing the "red/blue" regions define the 2DEG regions as determined by TCAD. The electron density $1.5 \times 10^{11} cm^{-2}$ is used as the criterion[1] for determining the 2DEG contours (black lines). In determining the 2DEG geometry, we apply the specified voltages for the TCAD analysis of the first device[19]. These values are: 25 volts on top gate (i.e., Al accumulation gate); -1.0 volt on depletion gates C12-C14; -1.0 volts on depletion gate C9 of Fig. 12; and zero volts on depletion gates C11 and C15. The TCAD simulation provides a new size and shape for the QD island C1 as shown in Fig. 12 (black line enclosing red and blue regions). The new size and shape for C1 is developed into the 3D solid model as shown in Fig. 13. The size and shape of the island dot C1 in Fig. 13 differs from that in Fig. 5 in the following way. The circular end aspects on the right and left are 11 nm closer to the poly side gates C11 and C15. The top portion of C1 is 7 nm farther away from poly side gates C12-C14. The bottom portion of C1 is 9 nm closer to the poly side gate C9, extending 4 nm underneath C9.

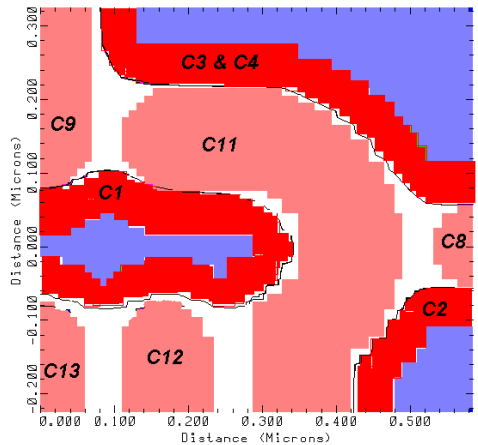

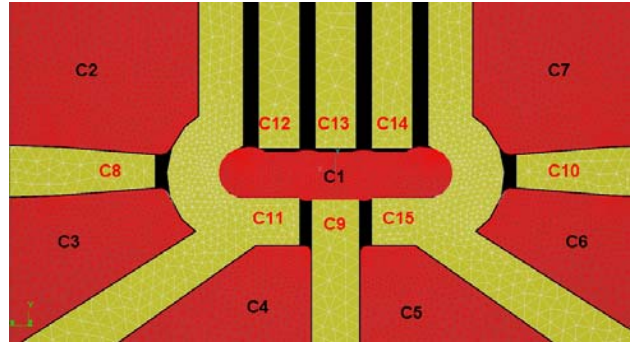

Fig. 12. Configuration of 2DEG conductors (red/blue areas) enclosed by a black line indicative of the $1.5 \times 10^{11}$ cm$^{-2}$ electron density contour), all overlaid with polysilicon conductors (rose areas), 35nm above silicon/$SiO_2$ interface, and top gate metal locations (blue areas) that extend down into the same plane as the polysilicon, 95nm above silicon/$SiO_2$ interface, Fig. 2).

Fig. 13. New size and shape for 2DEG position within the QD island, region C1, and outside, regions C2-C7, overlaid with the position of the poly-silicon conductor regions, C8-C15.

For the 3D solid model based on TCAD's 2DEG conductor C1 (i.e., Model 2), the resulting capacitances are given in Table 4. The only change between Model 1 and Model 2 is the shape and size of C1.

Table 4. Capacitance matrix for the quantum dot device[19] based on the TCAD estimated quantum dot geometry.

|    | 1       | 2       | 3       | 4       | 5       | 6       | 7       | 8       | 9       | 10      | 11      | 12      | 13      | 14      | 15      | 16      |
|----|---------|---------|---------|---------|---------|---------|---------|---------|---------|---------|---------|---------|---------|---------|---------|---------|
| 1  | 237.96  | -13.78  | -8.37   | -15.23  | -15.18  | -8.34   | -13.72  | -0.52   | -16.64  | -0.52   | -40.24  | -16.39  | -16.90  | -16.35  | -40.09  | -15.66  |
| 2  | -13.78  | 516.36  | -58.93  | -12.89  | -5.57   | -4.36   | -9.47   | -62.74  | -0.79   | -0.31   | -69.26  | -11.50  | -5.73   | -3.30   | -2.65   | -255.18 |
| 3  | -8.37   | -58.93  | 408.35  | -40.05  | -4.49   | -2.67   | -4.35   | -61.88  | -0.73   | -0.17   | -97.34  | -1.72   | -1.37   | -1.02   | -1.24   | -123.84 |
| 4  | -15.23  | -12.89  | -40.05  | 321.49  | -25.55  | -4.49   | -5.56   | -0.93   | -40.45  | -0.26   | -82.21  | -1.49   | -1.40   | -1.14   | -2.25   | -87.82  |
| 5  | -15.18  | -5.57   | -4.49   | -25.55  | 321.75  | -40.11  | -12.91  | -0.26   | -40.23  | -0.93   | -2.26   | -1.14   | -1.41   | -1.49   | -82.03  | -87.94  |
| 6  | -8.34   | -4.36   | -2.67   | -4.49   | -40.11  | 408.18  | -58.90  | -0.17   | -0.73   | -62.48  | -1.24   | -1.02   | -1.37   | -1.72   | -96.85  | -123.78 |
| 7  | -13.72  | -9.47   | -4.35   | -5.56   | -12.91  | -58.90  | 516.72  | -0.31   | -0.79   | -62.92  | -2.65   | -3.30   | -5.73   | -11.46  | -69.88  | -254.78 |
| 8  | -0.52   | -62.74  | -61.88  | -0.93   | -0.26   | -0.17   | -0.31   | 593.75  | -0.04   | -0.01   | -39.62  | -0.15   | -0.11   | -0.08   | -0.09   | -426.04 |
| 9  | -16.64  | -0.79   | -0.73   | -40.45  | -40.23  | -0.73   | -0.79   | -0.04   | 524.15  | -0.04   | -55.83  | -0.31   | -0.77   | -0.30   | -54.63  | -311.93 |
| 10 | -0.52   | -0.31   | -0.17   | -0.26   | -0.93   | -62.48  | -62.92  | -0.01   | -0.04   | 593.10  | -0.08   | -0.08   | -0.11   | -0.15   | -39.61  | -424.65 |
| 11 | -40.24  | -69.26  | -97.34  | -82.21  | -2.26   | -1.24   | -2.65   | -39.62  | -55.83  | -0.08   | 1499.23 | -177.34 | -3.06   | -1.26   | -0.92   | -925.63 |
| 12 | -16.39  | -11.50  | -1.72   | -1.49   | -1.14   | -1.02   | -3.30   | -0.15   | -0.31   | -0.04   | -177.34 | 534.18  | -175.86 | -3.00   | -1.27   | -138.81 |
| 13 | -16.90  | -5.73   | -1.37   | -1.40   | -1.41   | -1.37   | -5.73   | -0.11   | -0.77   | -0.11   | -3.06   | -175.86 | 533.13  | -176.15 | -3.05   | -140.69 |
| 14 | -16.35  | -3.30   | -1.02   | -1.14   | -1.49   | -1.72   | -11.46  | -0.08   | -0.30   | -0.15   | -1.26   | -3.00   | -176.15 | 531.73  | -175.41 | -138.52 |
| 15 | -40.09  | -2.65   | -1.24   | -2.25   | -82.03  | -96.85  | -69.88  | -0.09   | -54.63  | -39.61  | -0.92   | -1.27   | -3.05   | -175.41 | 1492.44 | -920.70 |
| 16 | -15.66  | -255.18 | -123.84 | -87.82  | -87.94  | -123.78 | -254.78 | -426.04 | -311.93 | -424.65 | -925.63 | -138.81 | -140.69 | -138.52 | -920.70 | 4375.98 |

The simulated capacitances between the depletion gates C9, C11-C15 and the quantum dot C1 are compared in Table 5, Model 2, with experimental results. Simulation agrees again within 10-20% and is in better agreement in many cases than using the heuristic approach, Model 1.

Table 5. Comparison of simulated and measured capacitances for the quantum dot device[18] based on TCAD estimated QD geometry.

|       | Capacitance | | |
|-------|-------------|----------------|----------|
| Gate  | Model 1     | Model 2 (TCAD) | Measured |
| C9,1  | 15 aF       | 16.7 aF        | 20.8 aF  |
| C11,1 | 36 aF       | 40.2 aF        | 42 aF    |
| C12,1 | 17 aF       | 16.4 aF        | 14.7 aF  |
| C13,1 | 17.5 aF     | 16.9 aF        | 11.2 aF  |
| C14,1 | 17 aF       | 16.4 aF        | 14.1 aF  |
| C15,1 | 36 aF       | 40.2 aF        | 42 aF    |

The Model 2 capacitances are sensitive to the density used in defining the edge of the dot C1. For example, if a density 10 time larger than the MIT critical density is used (i.e., $1.5 \times 10^{12}$ electrons/cm$^2$), the capacitances from the poly gates C9, C11-C15 to the dot C1 decrease by about 50% as the boundaries defining the dot's edge shrank by about 50 nm.

Agreement between model and experiment is highly dependent on the resultant process dimensions. The top gate capacitance to the quantum dot C16,1 as shown in Table 4 is 15.7 aF. Using parallel plate theory on the same TCAD estimated QD geometry, the top gate capacitance to the quantum dot is calculated to be 26.1 aF, which is 67% higher than the 3D capacitance value in Table 4. The 3D capacitance matrix was recalculated for a 10% change in the nominal 60 nm thickness of the $Al_2O_3$ layer (i.e., decreased to 54 nm and increased to 66 nm), while keeping all other aspects of the model fixed. The top gate capacitance to the quantum dot for the 54 nm $Al_2O_3$ layer thickness was computed to be 17.6 aF, an increase in capacitance of 12.5%, and for the 66 nm thickness, it was 13.7 aF, a decrease of 12.4%.

The capacitances from the poly gates C9, C11-C15 and the top gate C16 to the dot C1 were sensitive to the thickness of the $SiO_2$ gate oxide layer. Decreasing the thickness 10% from the nominal 35nm thickness resulted in those capacitances increasing by about 10%.

Though not treated here, another factor that to which the capacitances between the poly gates C9, C11-C15 to the dot C1 would be sensitive is the geometry of the poly gates (e.g., width dimension). The nominal width dimension of the poly gates C9, C11-C15 in Models 1 and 2 is 120-140 nm. If the geometry of the poly gates in models 1 and 2 is changed by 10%, the boundary edges would change by an offset distance of 6 nm (i.e., half of the total change in width). The resulting change in the distance from the poly gates to the dot would result in a significant change in the capacitances.

## 6. Discussion

The size and position of the quantum dot are important properties to establish from experimentally determined capacitances between the quantum dot and the neighboring gates. Few detailed analyses of the QD capacitances exist in the literature that consider the full three dimensional representation of the device structure and the impact of inaccuracies in the geometry on the resulting capacitances. Despite the challenges related to accurately calculating capacitances in a nanostructure with significant topography, the simulated capacitances agree with experiment within 10-20% in both the heuristic and TCAD approaches. The methodology shows similar agreement when describing charge sensing, discussed elsewhere[25].

The sensitivity of the capacitances to slight variations of the device geometry from nominal dimensions indicates that the disagreement between simulation and measurement is within the uncertainty produced by fabrication. For example, 10% increase or decrease in thickness of the $Al_2O_3$ layer from its nominal thickness produced a shift in the top gate capacitance to the quantum dot greater than 12%. Furthermore, a change of 10-20% in the width dimensions of the poly gates C9, C12-C15 will result in those capacitances changing by significant amounts. More accurate estimates of quantum dot capacitances will, therefore, be challenging without significant controls on the fabrication process.

The capacitance modeling is invaluable in helping to determine the position and size of the metallic quantum dot as it unambiguously indicates that the lithographic gate features define the dot. Disorder is known to often locally confine the electron density in much smaller regions than the intended lithographic structure. In the Si MOS system, this is a particular challenge especially in open lateral geometries [18]. The full 3D modeling of this system provides a clear picture of a metallic quantum dot island that is defined by the conducting gates in both structures and provides a critical analysis tool for experiment in determining position and size of the dots relative to their external gates.

The capacitance model agreement to experiment when using the metal-insulator transition density is a result that requires further investigation. It is not immediately clear why this choice of density should work so well because the electron wavefunction (or classically calculate density) extends beyond the MIT contour. Modulation of the gate potential should therefore modulate the integrated effective charge in this insulating region, which would manifest itself as additional capacitance and error. However, in the case of the lateral quantum dot, it appears that the density falls off is sufficiently rapidly that this is not a dominant source of error and that simultaneously the choice of this particular MIT density appears to be a relatively good one for identifying a metal edge that emulates the experimentally observed capacitances. Further investigation is necessary to determine whether using the MIT critical density was a fortunate first choice or if this is a more general result and for what regimes of the quantum dot it will remain a good estimate (e.g., metallic dots in contrast with few electron dots).

## 7. Conclusion

We have investigated the accuracy of two approaches to simulating the position and size of silicon quantum dot structures with significant topographical complexity. A three dimensional (3D) capacitance matrix is calculated for two significantly different quantum dot geometries in this work and show agreement within 10-20% of their respective experimental results published elsewhere. Uncertainty in the actual experimental structure due to small variations in layer thickness and gate sizes are more than sufficient to explain differences between the simulation and experimentally obtained capacitances. Furthermore, we present the observation that the size of the quantum dot, the metallic quantum confined region, can be estimated well using an electron density contour of $1.5 \times 10^{11}$ cm$^{-2}$ combined with semi-classical TCAD calculation of the electron density in sub-sections of the entire device structure. The metal-insulator transition for similar silicon MOS devices was measured recently[1] to be approximately $1.5 \times 10^{11}$ cm$^{-2}$. This model's utility is also demonstrated through its application in corroborating that the quantum dot size is defined by the lithographic gate features rather than disorder in the local potential. This model has potential applications both for assistance in analysis of complex quantum dot geometries, design of charge sensing, as well as evaluating behavior of future quantum-dot to circuit coupled configurations.

## 8. Acknowledgements

The authors are especially grateful for discussions with Dr. N. Zimmerman of the National Institute of Science and Technology regarding discussions about the gated nanowire MOS quantum dot structure and the application of the model to that structure. The authors are also grateful for discussions with Dr. Thomas M. Gurrieri of Sandia National Laboratories about the implications of capacitance modeling for charge sensing in double top gated geometries in relationship to future coupling to read-out circuitry. This work was supported by the Laboratory Directed Research and Development program at Sandia National Laboratories. Sandia is a multiprogram laboratory operated by Sandia Corporation, a Lockheed Martin Company, for the United States Department of Energy's National Nuclear Security Administration under Contract DE-AC04-94AL85000.**References**

[1] L. A. Tracy, E. H. Hwang, K. Eng, G. A. Ten Eyck, E. P. Nordberg, K. Childs, M. S. Carroll, M. P. Lilly, and S. Das Sarma, *"The observation of percolation-induced two-dimensional metal-insulator transition in a Si MOSFET"*, Physical Review B 79, 235307 (2009).

[2] Jan De Blauwe, "Nanocrystal nonvolatile memory devices," IEEE Transactions on Nanotechnology, Vol. 1, No. 1, March 2002, pp. 72-77.

[3] Wei Wu, Jian Gu, Haixiong Ge, Christopher Keimel, and Stephen Y. Chou, "Room-temperature Si single-electron memory fabricated by nanoimprint lithography," Applied Physics Letters 83, No. 11, 15 September 2003, pp. 2268-2270.

[4] M. Mitic, M. C. Cassidy, K. D. Petersson, R. P. Starrett, E. Gauja, R. Brenner, C. Yang, and D. N. Jamieson, "Demonstration of a silicon-based quantum cellular automata cell," Applied Physics Letters 89, 013503 (2006).

[5] Daniel Loss and David P. DiVincenzo, "Quantum computation with quantum dots," Physical Review A, Vol. 57, No. 1, January 1998, pp. 120-126.

[6] Yukinori Ono, Neil M. Zimmerman, Kenji Yamazaki and Yasuo Takahashi, "Turnstile operation using a silicon dual-gate single-electron transistor," Jpn. J. Appl. Phys. Vol. 42 (2003) pp. L1109-L1111, Part 2, No. 10A, 1 October 2003.

[7] L. J. Geerligs, V. F. Anderegg, P. A. M. Holweg, and J. E. Mooij, H. Pothier, D. Esteve, C. Urbina, and M. H. Devoret, "Frequency-locked turnstile device for single electrons," Physical Review Letters, Vol. 64, No. 22, 28 May 1990, pp. 2691-2694.

[8] R. J. Schoelkopf, P. Wahlgren, A. A. Kozhevnikov, P. Delsing, D. E. Prober, "The radio-frequency single-electron transistor (RF-SET): a fast and ultrasensitive electrometer," Science 280, 1238 (1998).
20